\documentclass[preprint]{aastex62}

\usepackage{amsmath,apjfonts,amsfonts,amssymb,enumerate,graphicx, longtable, float, times,textcomp,verbatim,url,nicefrac}
\usepackage[utf8]{inputenc}

\def\apjl{ApJ}%
\def\apjs{ApJS}%
\def\ao{Appl.~Opt.}%
\def\aap{A\&A}%
\newcommand{\flashpeakmag}{$14.49\pm 0.05$}

\newcommand{\afterglowpeakmag}{$17.65\pm 0.06$}

\newcommand{\trise}{$0.057 \pm 0.007$\ days} 
\newcommand{\tninety}{$0.5\pm 0.1$\ days}

\newcommand{\tdecay}{$0.23 \pm 0.03$\ days}

\newcommand{\tbreak}{$ 0.09 \pm 0.02$\ days after the {\it Fermi} trigger}
\newcommand{\alphaone}{$0.6 \pm 0.2$}
\newcommand{\alphatwo}{$-0.21 \pm 0.07$}
\newcommand{\correctedprompt}{13.40}

\tabletypesize{\footnotesize}
\shorttitle{Observations of GRB 230307A by TESS}
 \shortauthors{Fausnaugh et al.}

\begin{document}

\title{Observations of GRB 230307A by TESS}

\author[0000-0002-9113-7162]{Michael~M.~Fausnaugh}
\affil{Department of Physics and Kavli Institute for Astrophysics and Space Research, Massachusetts Institute of Technology, Cambridge, MA 02139, USA}
\correspondingauthor{Michael~M.~Fausnaugh}
\email{faus@mit.edu}

\author[0000-0002-7778-3117]{Rahul~Jayaraman}
\affil{Department of Physics and Kavli Institute for Astrophysics and Space Research, Massachusetts Institute of Technology, Cambridge, MA 02139, USA}

\author{Roland~Vanderspek}
\affil{Department of Physics and Kavli Institute for Astrophysics and Space Research, Massachusetts Institute of Technology, Cambridge, MA 02139, USA}

\author{George~R.~Ricker}
\affil{Department of Physics and Kavli Institute for Astrophysics and Space Research, Massachusetts Institute of Technology, Cambridge, MA 02139, USA}

\author[0000-0002-7754-9486]{Christopher~J.~Burke}
\affil{Department of Physics and Kavli Institute for Astrophysics and Space Research, Massachusetts Institute of Technology, Cambridge, MA 02139, USA}

\author[0000-0001-8020-7121]{Knicole D. Col\'{o}n}
\affiliation{NASA Goddard Space Flight Center, 8800 Greenbelt Rd, Greenbelt, MD 20771, USA}  

\author[0000-0003-0556-027X]{Scott W. Fleming}
\affil{Space Telescope Science Institute, 3700 San Martin Dr, Baltimore, MD 21218, USA}

\author[0000-0002-7871-085X]{Hannah M. Lewis}
\affil{Space Telescope Science Institute, 3700 San Martin Dr, Baltimore, MD 21218, USA}

\author[ 0000-0001-7106-4683]{Susan Mullally}
\affil{Space Telescope Science Institute, 3700 San Martin Dr, Baltimore, MD 21218, USA}


\author[0000-0002-1176-3391]{Allison Youngblood}
\affiliation{NASA Goddard Space Flight Center, 8800 Greenbelt Rd, Greenbelt, MD 20771, USA}

\author[0000-0001-7139-2724]{Thomas Barclay}
\affiliation{University of Maryland, Baltimore County, 1000 Hilltop Circle, Baltimore, MD 21250, USA}
\affiliation{NASA Goddard Space Flight Center, 8800 Greenbelt Rd, Greenbelt, MD 20771, USA}

\author[0000-0002-2942-3379]{Eric Burns}
\affiliation{Department of Physics and Astronomy, Louisiana State University, Baton Rouge, LA 70803 USA}

\author{David W.\ Latham}
\affil{Center for Astrophysics | Harvard \& Smithsonian, 60 Garden Street, Cambridge, MA 02138}

\author[0000-0002-6892-6948]{S.~Seager}
\affil{Department of Physics and Kavli Institute for Astrophysics and Space Research, Massachusetts Institute of Technology, Cambridge, MA 02139, USA}
\affil{Department of Earth, Atmospheric, and Planetary Sciences, Massachusetts Institute of Technology, Cambridge, MA 02139, USA}
\affiliation{Department of Aeronautics and Astronautics, MIT, 77 Massachusetts Avenue, Cambridge, MA 02139, USA}

\author[0000-0002-4265-047X]{Joshua~N.~Winn}
\affil{Department of Astrophysical Sciences, Princeton University, 4 Ivy Lane, Princeton, NJ 08544, USA}

\author{Jon M.\ Jenkins}
\affil{NASA Ames Research Center, Moffett Field, CA, 94035, USA}

\begin{abstract}
We present the TESS light curve of GRB 230307A.  We find two distinct components: a bright, prompt optical component at the time of the {\it Fermi} observation that peaked at TESS magnitude \flashpeakmag\ (averaged over 200 seconds), followed by a gradual rise and fall over 0.5 days, likely associated with the afterglow, that peaked at \afterglowpeakmag\ mag. The prompt component is observed in a single 200s Full Frame Image and was undetectable in the next TESS image ($T_{\rm mag} > 17.79$). Assuming that the onset of the optical transient was coincident with the gamma-ray emission, the prompt emission lasted less than 73.6 seconds, which implies the true peak was actually brighter than $T_{\rm mag} =$~\correctedprompt.  We also fit parametric models to the afterglow to  characterize its shape.  
The TESS TICA light curve can be retrieved at \url{https://tess.mit.edu/public/tesstransients/light_curves/lc_grb230307A_cleaned.txt}. 
\end{abstract}

\section{Introduction \label{sec:intro}}

GRB 230307A was a long gamma-ray burst discovered by {\it Fermi} at BJD$-2,460,000$=11.15549 days (15:44:06 UT on 2023 March 7, GCN33405; \citealt{GCN33405}).  
A fading optical afterglow was observed 9--56 hours after the GRB trigger by ULTRACAM (\citealt{GCN33439}), GMOS-S on Gemini South (\citealt{GCN33447}),  RASA36, and KMTnet (\citealt{GCN33449}).  The coordinates reported by \citet{GCN33439} are  RA = 04:03:25.83, Dec = $-$75:22:42.7.


The Transiting Exoplanet Survey Satellite (TESS, \citealt{Ricker2015}) observed the optical afterglow of GRB 230307A in the Full Frame Images (FFIs), sampled at 200 seconds. 
In this research note, we present the TESS light curve and observed properties of the optical emission.


\section{Observations and Data Reduction}

TESS observed GRB 230307A as part of its normal operations in Sector~62, which lasted from 2023 February 12 to March 10. The FFI data from March 3--10 were downloaded from the spacecraft on 2023 March 10, processed by the Payload Operations Center (POC) at MIT, and delivered to the Mikulski Archive for Space Telescopes (MAST) as a TICA High Level Science Product, \dataset[doi:10.17909/t9-9j8c-7d30]{https://doi.org/10.17909/t9-9j8c-7d30} \citep{Fausnaugh2020}.\footnote{\url{https://archive.stsci.edu/hlsp/tica}} The order of processing events was
\begin{itemize}
    \item March 10, 16:23~UTC---Deep Space Network (DSN) begins contact with the TESS spacecraft to downlink the data.
    \item March 11, 10:56~UTC---Compressed data arrive at POC from the DSN; POC begins decoding data delivery and writing image files.
    \item March 11, 16:35~UTC---POC begins calibrating data and fitting World Coordinate Solutions as part of the TICA data delivery.
    \item March 11, 18:42~UTC---POC begins staging and verifying the TICA delivery to MAST.
    \item March 12, 1:50~UTC---MAST begins data ingest.
        \item March 12, 6:40~UTC---GRB data are made public on MAST, announced on the MAST holdings page,\footnote{\url{https://outerspace.stsci.edu/display/TESS/TESS+Holdings+Available+by+MAST+Service}} and announced on MAST social media.
\end{itemize}

Once the data were public, we processed the images with the difference imaging analysis pipeline described by \citet{Fausnaugh2021}.  

We identified a transient point source in the TESS difference image overlapping with the {\it Fermi} trigger at BJD$-2,460,000$=11.15518 days, close to the coordinates reported by \citet{GCN33439}. We measured the position of the source using flux-weighted centroids in a 5x5 square pixel aperture. We found a position of RA = 04:03:25.36, Dec = $-$75:22:41.31, which agrees with the position from ULTRACAM to 1.8 arcseconds.  This difference is consistent with the typical 1$\sigma$ precision of TICA World Coordinate Solutions, about 2 arcseconds. 

\begin{figure*}
\centering
\includegraphics[width=\textwidth]{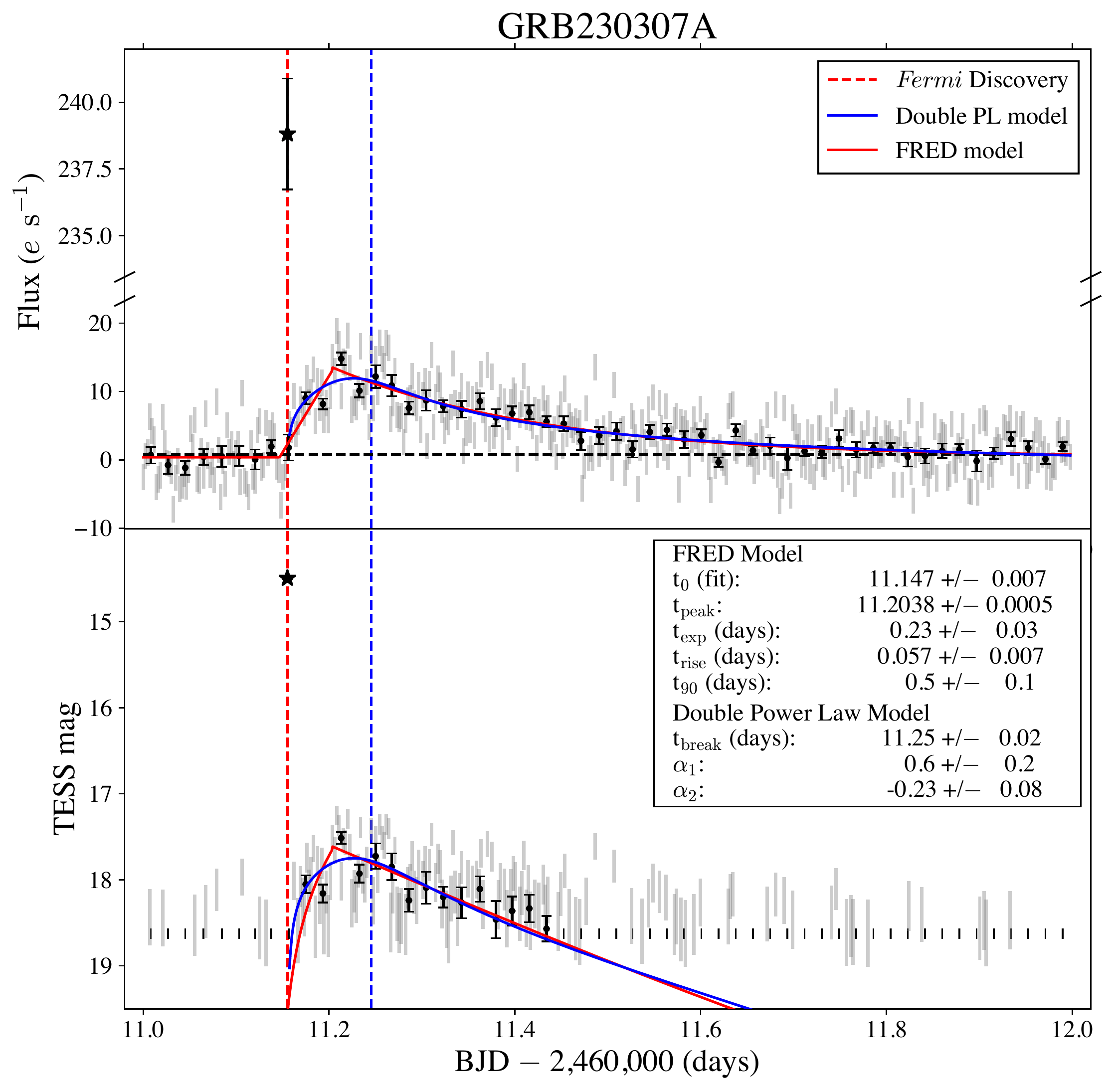}
\caption{\label{fig:grb_lc}TESS light curve of GRB 230307A.  The large star shows the prompt emission coincident with the {\it Fermi} discovery at BJD$-2,460,000$=11.15549 days (vertical red line).  The grey marks show the 200 second cadence light curve, and the black points show the light curve binned to 1600 seconds. The limiting magnitude in 1600 seconds is 18.63, based on the scatter of the light curve prior to the GRB discovery. All timestamps are given relative to 2,460,000 BJD.  Fits for two models of the afterglow are shown: a double power law (PL) and a fast-rise exponential-decay (FRED) model. For both fits, we excluded the prompt emission at the time of the {\it Fermi} trigger. Models were fit to the original light curve sampled at 200 seconds---data are binned here only for display purposes.  The horizontal dashed line shows the residual background of the FRED model, which is a nuisance parameter in our fits.  The vertical blue line shows the break time $t_{\rm break}$\ of the double power law.  The TESS light curve is available at \url{https://tess.mit.edu/public/tesstransients/light_curves/lc_grb230307A_cleaned.txt}.}
\end{figure*}

We extracted light curves from the difference images using forced PSF photometry at the location of the afterglow reported by \citet{GCN33439}.  Figure~\ref{fig:grb_lc} shows the resulting light curve. Based on the scatter of the light curve prior to the GRB discovery, we estimate a 5$\sigma$ limiting magnitude of 18.63 in 1600 seconds. The TESS light curve is available at \url{https://tess.mit.edu/public/tesstransients/light_curves/lc_grb230307A_cleaned.txt}. Information on the file format and processing steps is available on the same website.\footnote{\url{https://tess.mit.edu/public/tesstransients/pages/readme.html}}

 The light curve shows two distinct components.  First is a prompt rise from zero flux to TESS mag $T_{\rm mag} =$\ \flashpeakmag\ in a single FFI at BJD$-2,460,000$~$= 11.15518$ days (mid exposure). The quoted uncertainty includes a statistical uncertainty of 0.02 mag, but is dominated by uncertainty in the TESS instrument absolute flux calibration \citep{TIH}. The source is not detected in the next FFI, with a 3$\sigma$ upper limit of 17.79 mag. The second emission component starts just after the prompt emission, and consists of a gradual rise and fall in the TESS light curve over 0.5 days, likely associated with the afterglow.


\section{Analysis of the Afterglow}

We fit two parametric models to the afterglow light curve, a fast-rise exponential-decay (FRED) model and a double power law model.  We fit the original data sampled at 200 seconds. We excluded the prompt emission at $T_{\rm mag} = 14.49$ from the fits to the afterglow emission. The reported uncertainties correspond to the 68\% confidence intervals of the fits. For both models, we fit a residual background error as a nuisance parameter, which is consistent with zero at $1\sigma$.

From the FRED model, we find that
\begin{itemize}
    \item the peak magnitude of the afterglow is \afterglowpeakmag,
    \item the rise-time $t_{\rm rise} =$\ \trise, 
    \item the time containing 90\% of the flux $t_{90} =$\ \tninety,
    \item and the exponential decay timescale $t_{\rm exp} = $\ \tdecay.
\end{itemize}

For the double power law model, we used the parameterization of \citet{Li2012}:

\begin{align}
    F = F_{02}\left[ \left( \frac{t}{t_{\rm break}} \right)^{\alpha_1 \omega} + \left( \frac{t}{t_{\rm break}} \right)^{\alpha_2 \omega} \right]^{-1/\omega}
\end{align}
where $F_{02}$ is the flux normalization at the break time $t_{\rm break}$, while $\alpha_1$, and $\alpha_2$ are slope parameters. We fixed the smoothing parameter $\omega$ to a value of 5.0.  The best-fit model yields
\begin{itemize}
    \item $t_{\rm break} = $\ \tbreak,
    \item $\alpha_1 = $\ \alphaone,
    \item and $\alpha_2 = $\ \alphatwo.
\end{itemize}

We also fit the TESS light curve with the deep $r$-band magnitudes reported by \citet{GCN33439} and \citet{GCN33447}, and found consistent results.

\section{Discussion}

The end time of the FFI with the prompt emission is BJD$-2,460,000$=11.15634 days, while the {\it Fermi} trigger is at BJD$-2,460,000$=11.15549 days. Assuming that the onset of the optical light curve is coincident with the onset of the gamma-ray emission observed by {\it Fermi}, the prompt emission observed by TESS lasted less than 73.6 seconds. A shorter duration implies a higher peak magnitude for the same fluence; we find  $T_{\rm mag} = $\correctedprompt\ mag for a 73.6 second burst, and the peak could be even brighter if the duration of the prompt emission is shorter.   For a given luminosity distance $D$, the isotropic monochromatic luminosity at $7839$\,\AA\ (the pivot wavelength of the TESS filter) would be greater than $9\times 10^{44}(D/1\,\,{\rm Gpc})^2$\ erg s$^{-1}$. This calculation is corrected for Galactic extinction at $7839$\,\AA\ assuming a \citealt*{Cardelli1989} extinction law and  $E(B-V) = 0.0758$\ mag from \citealt{Schlafly2011}, but does not take into account the unknown spectral shape of the GRB in the broad TESS filter (6,000--10,000\,\AA).




These data demonstrate the utility of wide-field, continuous monitoring for studies of fast extragalactic transients.  TESS data released approximately every 7 days as TICA High Level Science Products will facilitate these studies over the next several years.


\section*{acknowledgments}
We thank the TESS-POC team at MIT and MAST for facilitating the TICA HLSP delivery. The TESS mission is funded by NASA’s Science Mission directorate.

\software{
  Matplotlib \citep{matplotlib},
  Numpy \citep{numpy},
  Scipy \citep{scipy}, 
  Astropy \citep{astropy},
  Astroquery \citep{astroquery},
  gwcs \citep{gwcs}, 
  TICA \citep{Fausnaugh2020}
 }

\bibliography{refs}

\begin{thebibliography}{}
\expandafter\ifx\csname natexlab\endcsname\relax\def\natexlab#1{#1}\fi

\bibitem[{{Astropy Collaboration} {et~al.}(2018){Astropy Collaboration},
  {Price-Whelan}, {Sip{\H o}cz}, {G{\"u}nther}, {Lim}, {Crawford}, {Conseil},
  {Shupe}, {Craig}, {Dencheva}, {Ginsburg}, {VanderPlas}, {Bradley},
  {P{\'e}rez-Su{\'a}rez}, {de Val-Borro}, {Aldcroft}, {Cruz}, {Robitaille},
  {Tollerud}, {Ardelean}, {Babej}, {Bach}, {Bachetti}, {Bakanov}, {Bamford},
  {Barentsen}, {Barmby}, {Baumbach}, {Berry}, {Biscani}, {Boquien}, {Bostroem},
  {Bouma}, {Brammer}, {Bray}, {Breytenbach}, {Buddelmeijer}, {Burke},
  {Calderone}, {Cano Rodr{\'{\i}}guez}, {Cara}, {Cardoso}, {Cheedella},
  {Copin}, {Corrales}, {Crichton}, {D'Avella}, {Deil}, {Depagne}, {Dietrich},
  {Donath}, {Droettboom}, {Earl}, {Erben}, {Fabbro}, {Ferreira}, {Finethy},
  {Fox}, {Garrison}, {Gibbons}, {Goldstein}, {Gommers}, {Greco}, {Greenfield},
  {Groener}, {Grollier}, {Hagen}, {Hirst}, {Homeier}, {Horton}, {Hosseinzadeh},
  {Hu}, {Hunkeler}, {Ivezi{\'c}}, {Jain}, {Jenness}, {Kanarek}, {Kendrew},
  {Kern}, {Kerzendorf}, {Khvalko}, {King}, {Kirkby}, {Kulkarni}, {Kumar},
  {Lee}, {Lenz}, {Littlefair}, {Ma}, {Macleod}, {Mastropietro}, {McCully},
  {Montagnac}, {Morris}, {Mueller}, {Mumford}, {Muna}, {Murphy}, {Nelson},
  {Nguyen}, {Ninan}, {N{\"o}the}, {Ogaz}, {Oh}, {Parejko}, {Parley}, {Pascual},
  {Patil}, {Patil}, {Plunkett}, {Prochaska}, {Rastogi}, {Reddy Janga},
  {Sabater}, {Sakurikar}, {Seifert}, {Sherbert}, {Sherwood-Taylor}, {Shih},
  {Sick}, {Silbiger}, {Singanamalla}, {Singer}, {Sladen}, {Sooley},
  {Sornarajah}, {Streicher}, {Teuben}, {Thomas}, {Tremblay}, {Turner},
  {Terr{\'o}n}, {van Kerkwijk}, {de la Vega}, {Watkins}, {Weaver}, {Whitmore},
  {Woillez}, {Zabalza}, \& {Astropy Contributors}}]{astropy}
{Astropy Collaboration}, {Price-Whelan}, A.~M., {Sip{\H o}cz}, B.~M., {et~al.}
  2018, \aj, 156, 123

\bibitem[{{Cardelli} {et~al.}(1989){Cardelli}, {Clayton}, \&
  {Mathis}}]{Cardelli1989}
{Cardelli}, J.~A., {Clayton}, G.~C., \& {Mathis}, J.~S. 1989, \apj, 345, 245

\bibitem[{Dencheva {et~al.}(2020)Dencheva, Mumford, Bradley, perrygreenfield,
  D'Avella, Sipőcz, Cara, Lim, Jones, Earl, Davies, Simon, Tollerud, Deil,
  Shanahan, Slavich, Streicher, Simpson, Hunkeler, Droettboom, \&
  de~Val-Borro}]{gwcs}
Dencheva, N., Mumford, S., Bradley, L., {et~al.} 2020, spacetelescope/gwcs:
  GWCS 0.15.0, doi:10.5281/zenodo.4271727

\bibitem[{{Fausnaugh} {et~al.}(2020){Fausnaugh}, {Burke}, {Ricker}, \&
  {Vanderspek}}]{Fausnaugh2020}
{Fausnaugh}, M.~M., {Burke}, C.~J., {Ricker}, G.~R., \& {Vanderspek}, R. 2020,
  Research Notes of the American Astronomical Society, 4, 251

\bibitem[{{Fausnaugh} {et~al.}(2021){Fausnaugh}, {Vallely}, {Kochanek},
  {Shappee}, {Stanek}, {Tucker}, {Ricker}, {Vanderspek}, {Latham}, {Seager},
  {Winn}, {Jenkins}, {Berta-Thompson}, {Daylan}, {Doty}, {F{\H{u}}r{\'e}sz},
  {Levine}, {Morris}, {P{\'a}l}, {Sha}, {Ting}, \& {Wohler}}]{Fausnaugh2021}
{Fausnaugh}, M.~M., {Vallely}, P.~J., {Kochanek}, C.~S., {et~al.} 2021, \apj,
  908, 51

\bibitem[{{Fermi GBM Team at MSFC/Fermi-GBM}(2023)}]{GCN33405}
{Fermi GBM Team at MSFC/Fermi-GBM}. 2023, GRB Coordinates Network, 33405, 1

\bibitem[{{Ginsburg} {et~al.}(2019){Ginsburg}, {Sip{\H o}cz}, {Brasseur},
  {Cowperthwaite}, {Craig}, {Deil}, {Guillochon}, {Guzman}, {Liedtke}, {Lian
  Lim}, {Lockhart}, {Mommert}, {Morris}, {Norman}, {Parikh}, {Persson},
  {Robitaille}, {Segovia}, {Singer}, {Tollerud}, {de Val-Borro}, {Valtchanov},
  {Woillez}, {The Astroquery collaboration}, \& {a subset of the astropy
  collaboration}}]{astroquery}
{Ginsburg}, A., {Sip{\H o}cz}, B.~M., {Brasseur}, C.~E., {et~al.} 2019, \aj,
  157, 98

\bibitem[{Hunter(2007)}]{matplotlib}
Hunter, J.~D. 2007, Computing in Science {\&} Engineering, 9, 90

\bibitem[{{Im} {et~al.}(2023){Im}, {Paek}, {Jeong}, {Chang}, {Choi}, \&
  {Lee}}]{GCN33449}
{Im}, M., {Paek}, G. S.~H., {Jeong}, M., {et~al.} 2023, GRB Coordinates
  Network, 33449, 1

\bibitem[{{Levan} {et~al.}(2023){Levan}, {Gompertz}, {Ackley}, {Kennedy}, \&
  {Dhillon}}]{GCN33439}
{Levan}, A.~J., {Gompertz}, B., {Ackley}, K., {Kennedy}, M., \& {Dhillon}, V.
  2023, GRB Coordinates Network, 33439, 1

\bibitem[{{Li} {et~al.}(2012){Li}, {Liang}, {Tang}, {Chen}, {Xi}, {L{\"u}},
  {Gao}, {Zhang}, {Zhang}, {Yi}, {Lu}, {L{\"u}}, \& {Wei}}]{Li2012}
{Li}, L., {Liang}, E.-W., {Tang}, Q.-W., {et~al.} 2012, \apj, 758, 27

\bibitem[{{O'Connor} {et~al.}(2023){O'Connor}, {Dichiara}, {Troja},
  {Gillanders}, {Cenko}, \& {Kouveliotou}}]{GCN33447}
{O'Connor}, B., {Dichiara}, S., {Troja}, E., {et~al.} 2023, GRB Coordinates
  Network, 33447, 1

\bibitem[{Oliphant(2007)}]{scipy}
Oliphant, T.~E. 2007, Computing in Science {\&} Engineering, 9, 10

\bibitem[{{Ricker} {et~al.}(2015){Ricker}, {Winn}, {Vanderspek}, {Latham},
  {Bakos}, {Bean}, {Berta-Thompson}, {Brown}, {Buchhave}, {Butler}, {Butler},
  {Chaplin}, {Charbonneau}, {Christensen-Dalsgaard}, {Clampin}, {Deming},
  {Doty}, {De Lee}, {Dressing}, {Dunham}, {Endl}, {Fressin}, {Ge}, {Henning},
  {Holman}, {Howard}, {Ida}, {Jenkins}, {Jernigan}, {Johnson}, {Kaltenegger},
  {Kawai}, {Kjeldsen}, {Laughlin}, {Levine}, {Lin}, {Lissauer}, {MacQueen},
  {Marcy}, {McCullough}, {Morton}, {Narita}, {Paegert}, {Palle}, {Pepe},
  {Pepper}, {Quirrenbach}, {Rinehart}, {Sasselov}, {Sato}, {Seager},
  {Sozzetti}, {Stassun}, {Sullivan}, {Szentgyorgyi}, {Torres}, {Udry}, \&
  {Villasenor}}]{Ricker2015}
{Ricker}, G.~R., {Winn}, J.~N., {Vanderspek}, R., {et~al.} 2015, Journal of
  Astronomical Telescopes, Instruments, and Systems, 1, 014003

\bibitem[{{Schlafly} \& {Finkbeiner}(2011)}]{Schlafly2011}
{Schlafly}, E.~F., \& {Finkbeiner}, D.~P. 2011, \apj, 737, 103

\bibitem[{van~der Walt {et~al.}(2011)van~der Walt, Colbert, \&
  Varoquaux}]{numpy}
van~der Walt, S., Colbert, S.~C., \& Varoquaux, G. 2011, Computing in Science
  {\&} Engineering, 13, 22

\bibitem[{{Vanderspek} {et~al.}(2018){Vanderspek}, {Doty}, Fausnaugh,
  Villase\~nor, {Jenkins}, {Berta-Thompson}, {Burke}, \& {Dreyer}}]{TIH}
{Vanderspek}, R., {Doty}, J., Fausnaugh, M., {et~al.} 2018,
  \href{https://archive.stsci.edu/missions/tess/doc/TESS_Instrument_Handbook_v0.1.pdf}{TESS
  Instrument Handbook}, Tech. rep., {Kavli Institute for Astrophysics and Space
  Science, Massachusetts Institute of Technology}

\end{thebibliography}

\end{document}